\documentclass[10pt, a4paper, twocolumn, showpacs, titlepage, showkeys, aps]{revtex4-1}
\usepackage{times}
\usepackage{amsmath}
\usepackage{amssymb}
\usepackage{graphicx}
\usepackage{subfigure}
\usepackage{booktabs}
\usepackage{longtable}
\usepackage{bm}
\usepackage{xcolor}
\usepackage{epsfig}
\usepackage{subfigure}
\usepackage{latexsym}
\usepackage{tabularx}
\usepackage{colortbl}
\usepackage{rotating}
\usepackage{comment}
\usepackage{natbib}
\usepackage{appendix}
\usepackage{amssymb, amsmath, amsbsy}

\begin{document}

\title{Exact solutions of the angular Teukolsky equation in particular cases}

\author{Chang-Yuan Chen$^1$}
\email{E-mail: chency@yctu.edu.cn, yctcccy@163.net}
\author{Yuan You$^1$}
\author{Xiao-Hua Wang$^1$}
\author{Fa-Lin Lu$^1$}
\author{Dong-Sheng Sun$^1$}
\author{Shi-Hai Dong$^2$}
\email{E-mail: dongsh2@yahoo.com.}
\affiliation{$^1$ School of Physics and Electronic Engineering, Yancheng Teachers University, Yancheng 224007, P.R.China\\
$^2$ Laboratorio de Informaci\'{o}n Cu\'{a}ntica, CIDETEC, Instituto Polit\'{e}cnico Nacional, UPALM, CDMX 07700, Mexico}
\pacs{03.67.-a, 03.65.Ud, 03.67.Mn}
\keywords{Angular Teukolsky equation, Exact solutions, Confluent Heun equation}

\begin{abstract}
In this work, we propose a new scheme to solve the angular Teukolsky equation for the particular case: $m=0, s=0$. We first transform this equation to a confluent Heun differential equation and then construct the Wronskian determinant to calculate the eigenvalues and normalized eigenfunctions. We find that the eigenvalues for larger $l$ are approximately given by  $_{0}{A_{l0}} \approx [l(l + 1) - \tau_{R}^2/2] - i\;\tau_{I}^2/2$ with an arbitrary $\tau^2=\tau_R^2 + i\,\tau_{I}^2$. The angular probability distribution (APD) for the ground state moves towards the north and south poles for $\tau_R^2>0$, but aggregates to the equator for $\tau_R^2\leq0$. However, we also notice that the APD for large angular momentum $l$ always moves towards the north and south poles , regardless the choice of $\tau^2$.

\end{abstract}
\maketitle

\section{Introduction}
The general form of the angular Teukolsky equation, also named as the spin-weighed spheroidal wave equation, has played an important role for studies of black holes with the gravitational self-force [1-4], quasi-normal modes [5-8], etc. Explicitly the equation is given in the form as [9, 10]
\begin{equation}\label{sst}
\begin{array}{c}
\displaystyle(1 - {x^2})\frac{{{d^2}{}_s{S_{lm}}(\tau , x)}}{{d{x^2}}} - 2x\frac{{d{}_s{S_{lm}}(\tau , x)}}{{dx}} + \Big[{}_s{A_{lm}}(\tau ) + s \\[2mm]
\displaystyle + {\tau ^2}{x^2} - 2s\, \tau\,  x - \frac{{{{(m + s\, x)}^2}}}{{(1 - {x^2})}} \Big]{}_s{S_{lm}}(\tau , x) = 0
\end{array}
\end{equation}
where $x=\cos \theta \in[-1, 1], \theta  \in[0, \pi]$ and $\tau=a\, \omega $. The parameter $a$  is angular momentum of per unit mass of the black holes,  $\omega$ is a complex frequency, $l = 0, \;1, \;2, \;3, \cdots$ are angular quantum number and  $m=0, \pm 1, \pm 2, \cdots, \pm l$ are magnetic quantum number.  The spin weight of the field $s$ is given by $s =\pm 2$ for gravitational perturbations, $s =\pm 1$ for electromagnetic perturbations, $s =\pm 1/2$ for massless neutrino perturbations, and  $s = 0$ for scalar perturbations. The eigenfunctions ${}_s{S_{lm}}(\tau, \pm 1)$ and eigenvalues ${}_s{A_{lm}}(\tau )$ are required to be bounded according to the natural boundary conditions.

If taking $s = 0, \tau ^2 =  - c^2$  in Eq.(1) (the choice of the sign before $c^2$ is different, e.g. a positive sign was used in Refs. [11-13]) then it becomes a well-known spheroidal wave equation [14, 15]
\begin{equation}\label{spheroidal}
\begin{array}{c}
\displaystyle(1 - {x^2})\frac{{{d^2}{}_0{S_{lm}}(c, x)}}{{d{x^2}}} - 2x\frac{{d{}_0{S_{lm}}(c, x)}}{{dx}} \\[2mm]
\displaystyle + \left[ {{}_0{A_{lm}}(c) - {c^2}{x^2} - \frac{{{m^2}}}{{(1 - {x^2})}}} \right]{}_0{S_{lm}}(c, x) = 0
\end{array}
\end{equation}
which has an important application in the electromagnetic theory, e.g., spheroidal wave functions are frequently encountered, especially when boundary value problems in spheroidal structures are solved using full-wave analysis [15].

So far the exact solutions of Eq. (1) have not been obtained except for two particular cases, i.e., 1) for  $s=0$ and $\tau=0$, and 2) for $\tau=0$, $s\ne0$. The solutions of first case are well known, i.e., its eigenvalues and eigenfunctions are given by $_0{A_{lm}}(0)=l(l + 1)$ and normalized associated Legendre polynomials ${}_0{S_{lm}}(0, x) = {N_{lm}}P_l^m(x)$ [16, 17]. The solutions of second case were first given in [18] with the eigenvalues $_s{A_{lm}}(0) = l(l + 1) - s(s + 1)$  and eigenfunctions $_{s}{S_{lm}}(0, x)={}_{s}P_{lm}(x)=N'_{lms}(1 - x)^{(m + s)/2}(1 + x)^{(m - s)/2}P_{\;\quad \quad n}^{(m + s, m - s)}(x)$. These were confirmed in our recent studies [19] and $P_{\;\;\;n}^{(\alpha , \beta )}(x)$  are Jacobi polynomials. The reason why one cannot obtain the exact solutions of Eq.(1) or (2) is from the term
$\tau^2\, x^2$ ($\tau$ is a complex number). Up to now, different approaches were used to study their solutions numerically or approximately [11-13, 20-26]. To show its role, we shall explore how to obtain the exact solutions of the following equation
\begin{equation}\label{}
\begin{array}{c}
\displaystyle (1 - {x^2})\frac{{{d^2}{}_0{S_{l0}}(\tau , x)}}{{d{x^2}}} - 2x\frac{{d{}_0{S_{l0}}(\tau , x)}}{{dx}} \\[2mm]
\displaystyle + \left[ {{}_0{A_{l0}}(\tau ){\rm{ + }}{\tau ^2}{x^2}} \right]{}_0{S_{l0}}(\tau , x) = 0.
\end{array}
\end{equation}
Obviously, it is the special case of Eq.(1) or (2) for $s=0, m=0$. The exact solutions of Eq. (3) are very important for solving Eq.(1) [27-32]. Here, we propose a new scheme to solve this equation (3).

The rest of this work is organized as follows. In Section II, we propose a new scheme to solve equation (3) and present its analytical solutions. In Section III, we evaluate the eigenvalues through solving the Wronskian determinant and illustrate the property of the normalization wave functions. Finally we summarize our conclusions in Section IV.

\section{Exact solutions}

Taking function transformation ${}_0{S_{l0}}(\tau , x) = {e^{\tau x}}f(x)$  and a new variable  $z = (1 - x)/2, (-1 \le x \le 1\, , \;0 \le z \le 1\, )$, equation (3) can be transformed to a confluent Heun differential equation
\begin{equation}\label{}
\begin{array}{c}
\displaystyle \frac{{{d^2}f(z)}}{{d{\kern 1pt} {z^2}}} + \left[ { - 4\tau  + \frac{1}{z} + \frac{1}{{z - 1}}} \right]\frac{{d{\kern 1pt} f(z)}}{{d{\kern 1pt} z}} \\[2mm]
\displaystyle + \left[ {\frac{{{}_0{A_{l0}}(\tau ) + {\tau ^2} - 2\tau }}{z} + \frac{{ - {}_0{A_{l0}}(\tau ) - {\tau ^2} - 2\tau }}{{z - 1}}} \right]f(z) = 0.
\end{array}
\end{equation}

Compared this with the standard form of the confluent Heun equation [33,34]
\begin{equation}\label{}
\begin{array}{c}
\displaystyle \frac{{{d^2} H(z)}}{{d{\kern 1pt} {z^2}}} + \left[ {\alpha  + \frac{{\beta  + 1}}{z} + \frac{{\gamma  + 1}}{{z - 1}}} \right] \frac{{d{\kern 1pt} H(z)}}{{d{\kern 1pt} z}} \\[2mm]
\displaystyle + \left[ {\frac{\mu }{z} + \frac{\nu }{{z - 1}}} \right]H(z) = 0
\end{array}
\end{equation}
we have
\begin{equation}\label{}
\begin{array}{l}
\alpha=-4\tau, \beta=\gamma=0, \mu={}_0{A_{l0}}(\tau)+{\tau^2}-2\tau, \\[2mm]
\nu=-{}_0{A_{l0}}(\tau)-{\tau ^2}- 2\tau , \delta= 0, \eta=-{}_0{A_{l0}}-\tau^2.
\end{array}
\end{equation}

Thus, the solutions of Eq.(5) can be expressed as
\begin{equation}\label{}
\begin{array}{l}
H(z) = {\rm{HeunC}}(\alpha , \beta , \gamma , \delta , \eta , z)\\[2mm]
~~~~~~~~~= \sum\limits_{n = 0}^\infty  {\upsilon _n} (\alpha , \beta , \gamma , \delta , \eta ){z^n}, z\in[0, 1),
\end{array}
\end{equation}
where  $\delta=\mu+\nu-[\alpha(\beta+\gamma+2)/2], \eta=[\alpha(\beta+1)/2]-[(\beta+\gamma+\beta\gamma)/2]-\mu $. When ${\rm{HeunC}}(\alpha, \beta, \gamma, \delta, \eta, 0)=1$, one has the recurrent relation ${A_n}{\upsilon _n} = {B_n}{\upsilon _{n - 1}} + {C_n}{\upsilon _{n - 2}}$ with initial coefficients $ {\upsilon _{ - 1}} = 0, {\upsilon _0} = 1$ . Under constraints: 1) ${\Delta _{N + 1}}(\mu ) = 0$, and 2) $ \mu  + \nu  + N\alpha  = 0$, the confluent Heun functions ${\rm{HeunC}}(\alpha , \beta , \gamma , \delta , \eta , z)$  shall be truncated to $N$-term polynomials [33, 34] and also satisfy the natural condition at $z=1$, i.e., the wave function is convergent and finite at this limit. Unfortunately, it is known from Eq.(6) that the second constraint 2) is violated. Thus, we may only obtain the convergent solution of Eq. (3) at $z=0$
(north pole $\theta=0, \;x=1$) as
\begin{equation}\label{}
Y(1)=S(x)=e^{\tau x}H(1)
\end{equation} where $S(x)=e^{\tau x}{\mathop{\rm HeunC}\nolimits} ( - 4\tau , 0, 0, 0, - {}_0{A_{l0}} - {\tau ^2}, (1 - x)/2)$ are non-normalized eigenfunctions. With the same transformation as above for ${}_0{S_{l0}}(\tau , x)$, but with $z' = (1 + x)/2$ , ($ 0 \le z' \le 1$ ), equation (3) can be transformed to another form of confluent Heun equation
\begin{equation}\label{}
\begin{array}{c}
\displaystyle \frac{{{d^2}f(z')}}{{d{\kern 1pt} {{z'}^2}}} + \left[ {4\tau  + \frac{1}{{z'}} + \frac{1}{{z' - 1}}} \right]\frac{{d{\kern 1pt} f(z')}}{{d{\kern 1pt} z'}} \\[2mm]
\displaystyle + \left[ {\frac{{{}_0{A_{l0}}(\tau ) + \tau^2 + 2\tau }}{{z'}} + \frac{{ - {}_0{A_{l0}}(\tau ) - \tau^2 + 2\tau }}{{z' - 1}}} \right]f(z') = 0
\end{array}
\end{equation}
from which we are able to obtain all corresponding parameters $\alpha, \beta, \gamma, \mu, \nu, \delta, \eta$ by replacing $\tau$ in Eq. (6) with $-\tau$.
Thus, we obtain the convergent solution of Eq. (3) at $z'=0$ (south pole $\theta=\pi, x =-1$  ) as
\begin{equation}\label{}
\begin{array}{l}
Y(2)=e^{\tau x}{\mathop{\rm HeunC}\nolimits} (4\tau , 0, 0, 0, - {}_0{A_{l0}} - \tau^2, (1 + x)/2)\\[2mm]
~~~~~~~~=e^{\tau x}H(2).
\end{array}
\end{equation}Likely, the second constraint 2) is still not satisfied.

\section{Wronskian determinant and property of wave functions}

Eqs.(8) and (10) as the solutions of Eq. (3) at north- and south- poles should be convergent for a correct and same eigenvalue ${}_0A{}_{l0}$ and they must also be linearly dependent within the interval $x\in (-1,1)$. Nevertheless, for non-zero constants $C_1$  and $C_2$, one has
$C_{1} Y(1)+ C_{2}Y(2)=0$. Substitution of Eqs. (8) and (10) into this equation allows us to obtain
$C_{1}H(1) + C_{2}H(2)=0$ and its first derivative $C_1H'(1) + C_{2}H'(2) = 0$, from which we obtain the Wronskian determinant
\begin{equation}\label{deter}
\left| {\begin{array}{*{20}{c}}
{H(1)}&{H(2)}\\
{H'(1)}&{H'(2)}
\end{array}} \right| = 0,
\end{equation}
Since two solutions (8) and (10) are linearly dependent in the whole interval $x\in(-1,1)$, we take $x=0$ to calculate (\ref{condition1}) for simplicity. That is, its explicit expression is given by
\begin{widetext}
\begin{equation}\label{condition1}
\begin{array}{l}
 \operatorname{HeunC}(-4\tau ,0 ,0 ,0 ,- {}_0{A_{l0}} - \tau^2 ,0.5)\operatorname{HeunC}\text{Prime}(4\tau ,0 ,0 ,0 ,- {}_0{A_{l0}} - \tau^2 ,0.5) \\
 \text{+}\operatorname{HeunC}(4\tau ,0 ,0 ,0 ,- {}_0{A_{l0}} - \tau^2 ,0.5)\operatorname{HeunC}\text{Prime}(-4\tau ,0 ,0 ,0 ,- {}_0{A_{l0}} - \tau^2 ,0.5)=0.
\end{array}
\end{equation}
\end{widetext}
Such a calculation can be performed by Maple.

When $\tau^2=\tau_R^2$  is a real number, the operator of boundary-value problem for Sturm-Liouville problem corresponding to Eq. (3)
\begin{equation}\label{}
L=-\frac{d}{{dx}}(1-{x^2})\frac{d}{{dx}}-\tau_R^2\, x^2
\end{equation}
is a Hermite operator (self - adjoint operator) [35,36]. Thus, its eigenvalues are necessarily real, and its eigenfunctions belonging to different eigenvalues are orthogonal to each other.
When $\tau^2=0$, the formula (\ref{condition1}) is reduced to $\operatorname{HeunC}(0 ,0 ,0 ,0 ,- {}_0{A_{l0}},0.5)\operatorname{HeunC}\text{Prime}(0,0 ,0 ,0 ,- {}_0{A_{l0}},0.5)=0$. Let ${}_0{A_{l0}}=l(l+1)$,
one has $\operatorname{HeunC}(0 ,0 ,0 ,0 ,- {}_0{A_{l0}},0.5)=0$ for $l=1,3,5,\ldots$, while $\operatorname{HeunC}\text{Prime}(0,0 ,0 ,0 ,- {}_0{A_{l0}},0.5)=0$ for $l=0,2,4,\ldots$. Therefore, when $l=0,1,2,3,4,\ldots$, the non normalized eigenfunctions are given by $_{0}S_{l0}(0,x)=\operatorname{HeunC}(0 ,0 ,0 ,0 ,-l(l+1),(1-x)/2)=\operatorname{HeunC}(0 ,0 ,0 ,0 ,-l(l+1),(1+x)/2)=P_{l}(x)$. When $\tau^2$  is positive real number
($\tau$ is real number too), it is known from Eqs.(8) and (10) that the $Y(1), Y(2), H(1), H(2)$ are all real functions. Therefore, the following function
\begin{equation}\label{difference}
F(\tau , A) = H(1)H'(2) - H'(1)H(2)\end{equation}
represents its variation to the eigenvalues ${}_0{A_{l0}}$ (abbreviated as $A$)  for some given real number $\tau$ . Its intersections with the axis of $A$ decide the eigenvalue ${}_0{A_{l0}}$.
The eigenvalues $_{0}A_{l0}$ for different angular momentum $l$ can be calculated precisely by solving Eq.(12), and results at different values of $\tau^2$ (i.e., $-5, 5$) are listed in the first and the last columns of Table I.
When $\tau^2<0$  is a negative real number, $\tau$ becomes an imaginary number. Thus, all functions $Y(1), Y(2), H(1), H(2)$ are complex and the function
\begin{equation}\label{}
{\mathop{\rm Re}\nolimits} [F(\tau , A)] = {\rm Re}[H(1)H'(2) - H'(1)H(2)]
\end{equation}
denotes its variation to $A$. Its intersections with the axis $A$ determines the eigenvalues ${}_0{A_{l0}}$, but the function ${\mathop{\rm Im}\nolimits} [F(\tau , A)] = {\rm Im} [H(1)H'(2) - H'(1)H(2)]$ is always zero.

As illustrated in Fig. 1(a), we plot the variation of $F(\tau , A)$  with respect to $A$ for three different cases $\tau  = 0\;, \;\sqrt 5 , \;i\sqrt 5$.
It is known from the first column and the last row in Table I as well as other calculations for real $\tau^2=\tau_{R}^2$ that we observe that the eigenvalues for a larger $l$ are given by
\begin{equation}\label{}
{}_0{A_{l0}} \approx l(l + 1)- \tau_R^2/2.
\end{equation}This coincides with the results of Refs. [12,13] as shown in Table II.

\begin{widetext}
\begin{table*}
\caption{The eigenvalues $A =A_{R} + i\;A_{I}$  for $\tau^2$ at a couple of values.} 
\begin{center}
\begin{tabular}{c | c | cc | cc | cc | c }
\hline
 $l$  & $\tau^2=-5$ & \multicolumn{2}{c|}{$\tau^2=-5-8i$} & \multicolumn{2}{c|}{$\tau^2=-8i$} & \multicolumn{2}{c|}{$\tau^2=5-8i$} & $\tau^2=5$ \\
 \raisebox{2.3ex}[0pt] & $A_R$ & $A_{R}$  & $A_{I}$ & $A_{R}$  & $A_{I}$ & $A_{R}$  & $A_{I}$ & $A_R$ \\
\hline
0   &  1.3573568373 &     1.8043359712 &  1.5906398710 &    0.9644232667 &  2.3283101734 &    -0.7871216011  & 4.2776554165 &    -2.0799341864 \\
1   &  4.8228091767 &     5.3060843684 &  4.2234037629 &    2.4498686232 &  4.8343804401 &    -0.7986839845  & 5.3524223669 &    -1.1624779006 \\
2   &  8.8107354528 &     8.6711184570 &  4.8951208631 &     5.3319345225 &  4.5322189468 &    2.6808507613  &  2.9479975338 &    3.6779585066 \\
3   &  14.643458488 &    14.389856477  &  4.3808340000 &    11.777888730 &  4.0551516290 &     9.3813263521  &  3.8210930766 &    9.5179821017 \\
4   &  22.577779187 &    22.455887633  &  4.1922370585 &    19.888508215 &  4.0487913571 &     17.400494074  &  3.9144653289 &    17.511597841 \\
5   &  32.549800782 &    32.476062073  &  4.1242211276 &    29.927930974 &  4.0336407119 &      27.435200980  & 3.9441485104 &     27.506765233 \\
6   &  44.534892992 &    44.484067980  &  4.0871691371 &    41.949663846 &  4.0240585160 &      39.454371034 &  3.9613662124 &     39.504498010 \\
7   &  58.525881046 &    58.488474447  &  4.0646308842 &    55.962779217 &  4.0180276814 &      53.466085136 &  3.9715910879 &     53.503221201 \\
8   &  74.519987827 &    74.491226293  &  4.0498876033 &    71.971320202 &  4.0140030147 &      69.473787227 &  3.9781931777 &     69.502428232 \\
9   &  92.515913447 &    92.493078225  &  4.0396999811 &    89.977204764 &  4.0111886530 &       87.479124307 &  3.9827143032 &     87.501900094 \\
10 &  112.51297542 &    112.49439092   & 4.0323582429 &    109.98143670 &  4.0091448755 &       107.48297683 &  3.9859512535 &     107.50152964 \\
\hline
\end{tabular}
\end{center}
\end{table*}
\end{widetext}

The normalized even- and odd- parity eigenfucntions for positive and negative real number $\tau^2$  are plotted by using the obtained eigenvalues. It is difficult to observe the hidden symmetry in Eq.(8), which is expressed by the combination of an exponential function and an infinite series, but such a symmetry is shown explicitly in Fig. 2.

When $\tau^2{\rm{ = }}\tau_R^2 \pm i\, \tau_{I}^2$  is a complex number, the corresponding operator for Eq. (3)
\begin{equation}\label{operator-2}
L_{\pm}=-\frac{d}{{dx}}(1 - {x^2})\frac{d}{{dx}} - (\tau_R^2 \pm i\;{\tau_I^2})x^2
\end{equation} is not a Hermitian any more. Hence, the eigenvalues can only take complex values $ {A_ \pm } = {A_R} \pm i\;{A_I}$  and eigenfunctions are complex $Y_{\pm}$ [35, 36]. It is easy to show that $L_- = L_{+}^*$ and $L_{+} = L_{-}^*$. This implies that $A_{-} = A_{+}^{*} =A_{R}-i\, A_{I}$  and $Y_{-}=Y_{+}^{*}$  . Based on (\ref{deter}) we may plot the following two functions
\begin{equation}\label{}
{\mathop{\rm Re}\nolimits} [F(\tau , A_{R} + i\, A_{I})] = {\mathop{\rm Re}\nolimits} [H(1)H'(2)-H'(1)H(2)] = 0
\end{equation}
\begin{equation}\label{}
\quad {\mathop{\rm Im}\nolimits} [F(\tau , A_{R} + i\, A_{I})] = {\mathop{\rm Im}\nolimits} [H(1)H'(2) - H'(1)H(2)] = 0
\end{equation}
Their intersections decide the real and imaginary parts of the eigenvalues. In Fig.1(b)and (c), we show the plots of $\tau^2 =-5-8i$  and $\tau^2=-5+ 8i$. We find that the values of the intersection are conjugated each other and list their results at different values of $\tau^2$ (i.e., $-5-8i$, $-8i$, $5-8i$) in the middle columns of Table I. If $\tau^2=\tau_R^2 + i\, \tau_{I}^2$, we observe that the eigenvalues for a larger $l$ follow
\begin{equation}\label{}
_{0}{A_{l0}} \approx [l(l + 1)- \tau_{R}^2/2]- i\;\tau_{I}^2/2.
\end{equation}

Since the eigenvalues and eigenfunctions for complex $\tau^2 = \tau_R^2 \pm \;i\;\tau_{I}^2$ are conjugated, we show in Fig. 3 the real and imaginary parts of the eigenfunctions for the case $\tau^2 =-5-8i$. We note that only if $\tau^2$  is complex, the real or imaginary parts of the eigenfunctions are odd or even functions. Its odd or even parity property coincides with that of the angular momentum $l$. This is the same as the case when $\tau^2$  is real. However, the number of nodes for complex $\tau^2$  is not equal to the value $l$, which is different from the case for real $\tau^2$, where the number of nodes is $l$.
\begin{figure}
\subfigure[]{\includegraphics[width=7.6cm]{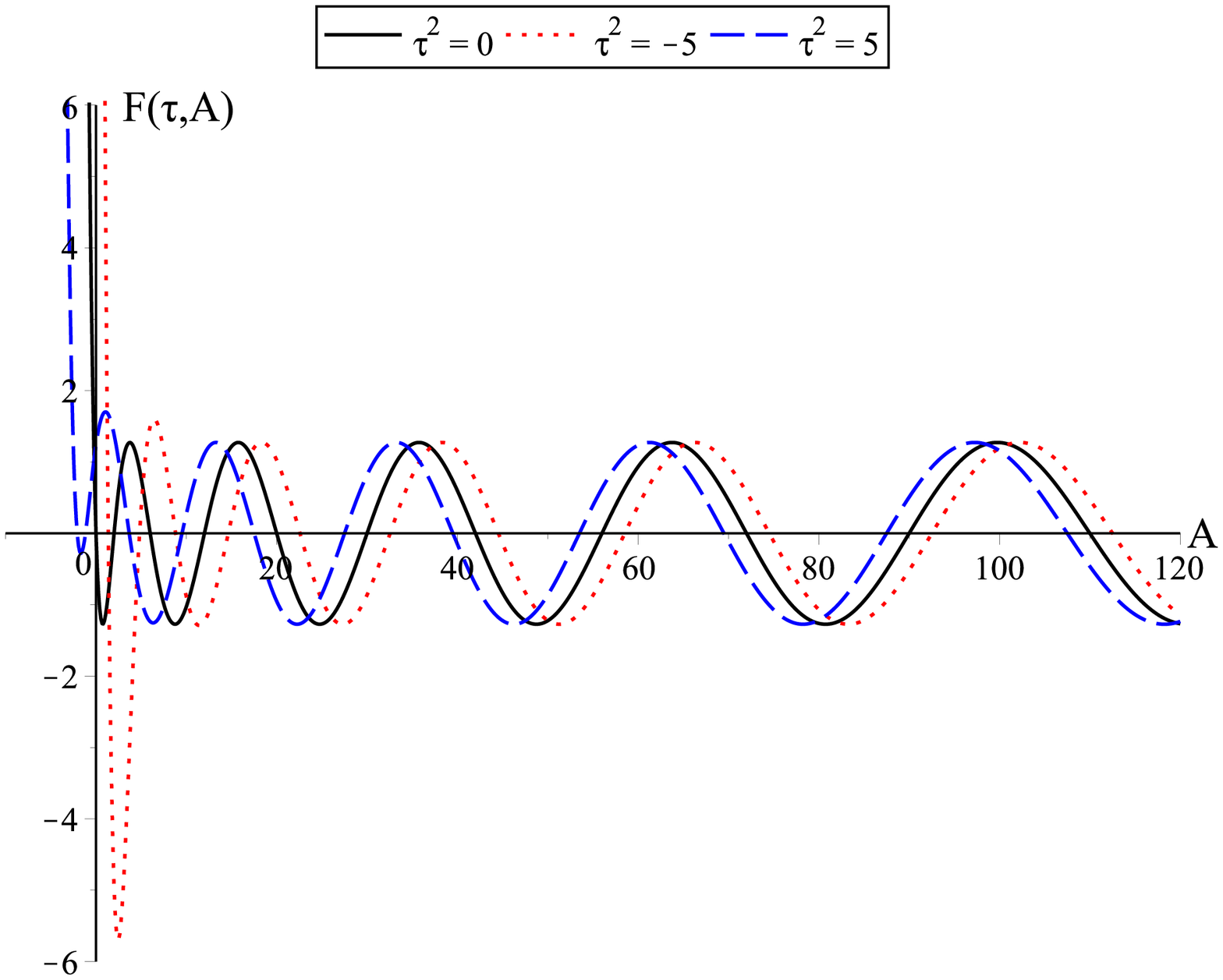}}\label{CA1}\hfil%
\subfigure[]{\includegraphics[height=3.5cm]{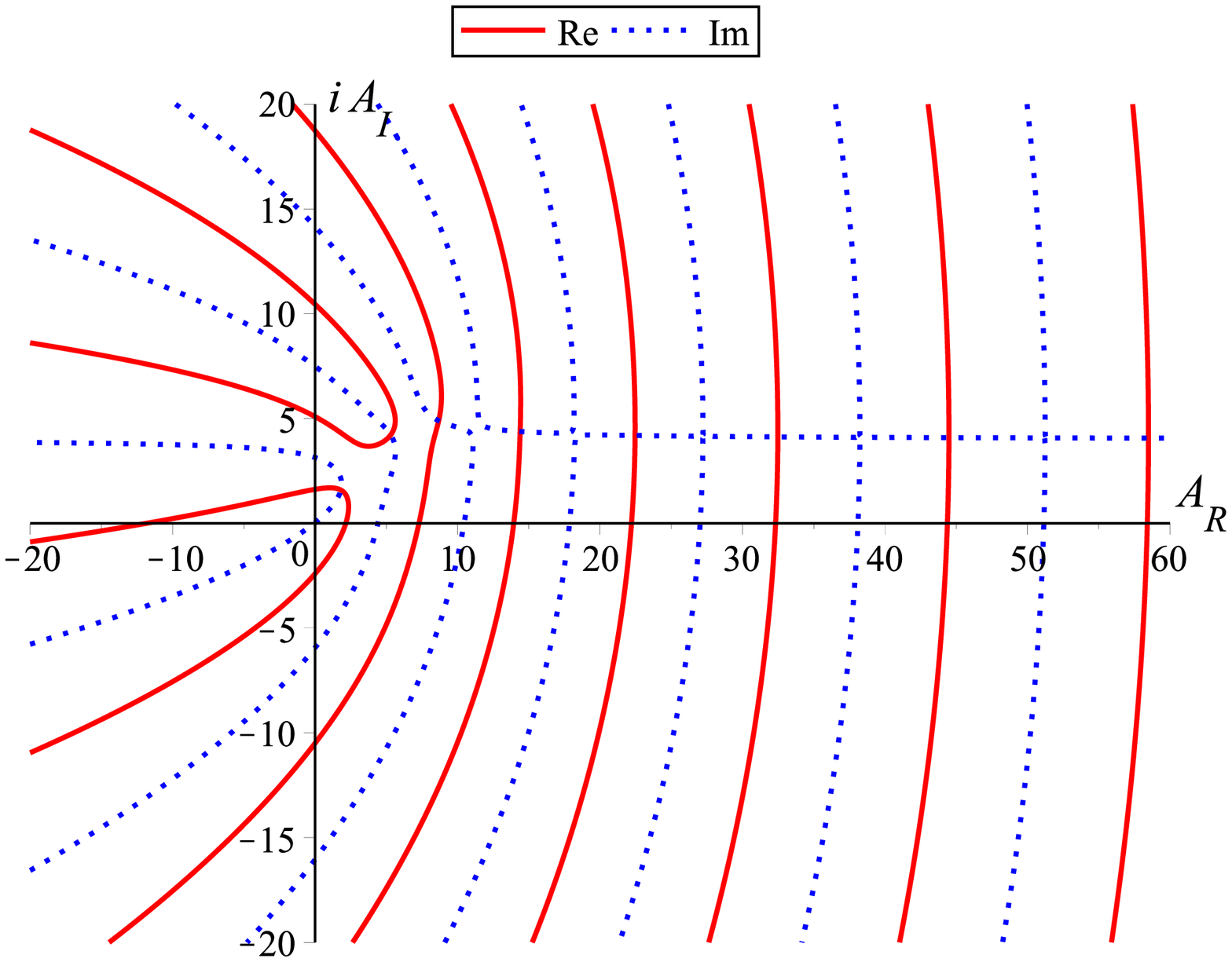}}\label{CA1}\hfil%
\subfigure[]{\includegraphics[height=3.5cm]{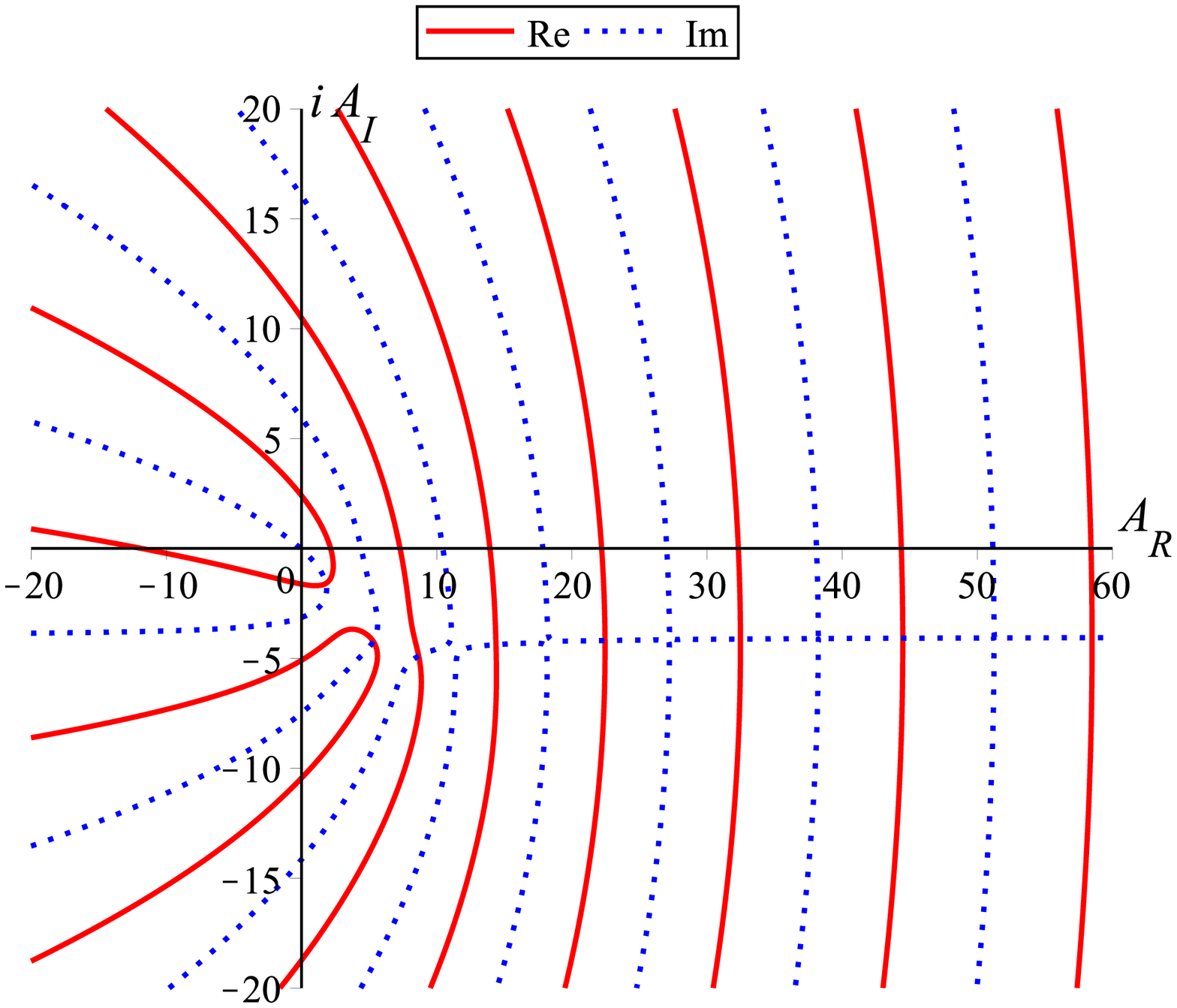}}\label{NA1}
\caption{The eigenvalues denotations in complex plane. a) The case $ \tau^2 = 0\;, \; \pm 5\;$  b)the case $\tau^2=-5-8i$ and c)$\tau^2=-5+8i$ }
\label{fig:condiciones}
\end{figure}

\begin{figure}
\label{fig::fw1}
\begin{tabular}{l}
\subfigure[]{\includegraphics[width=3.8cm]{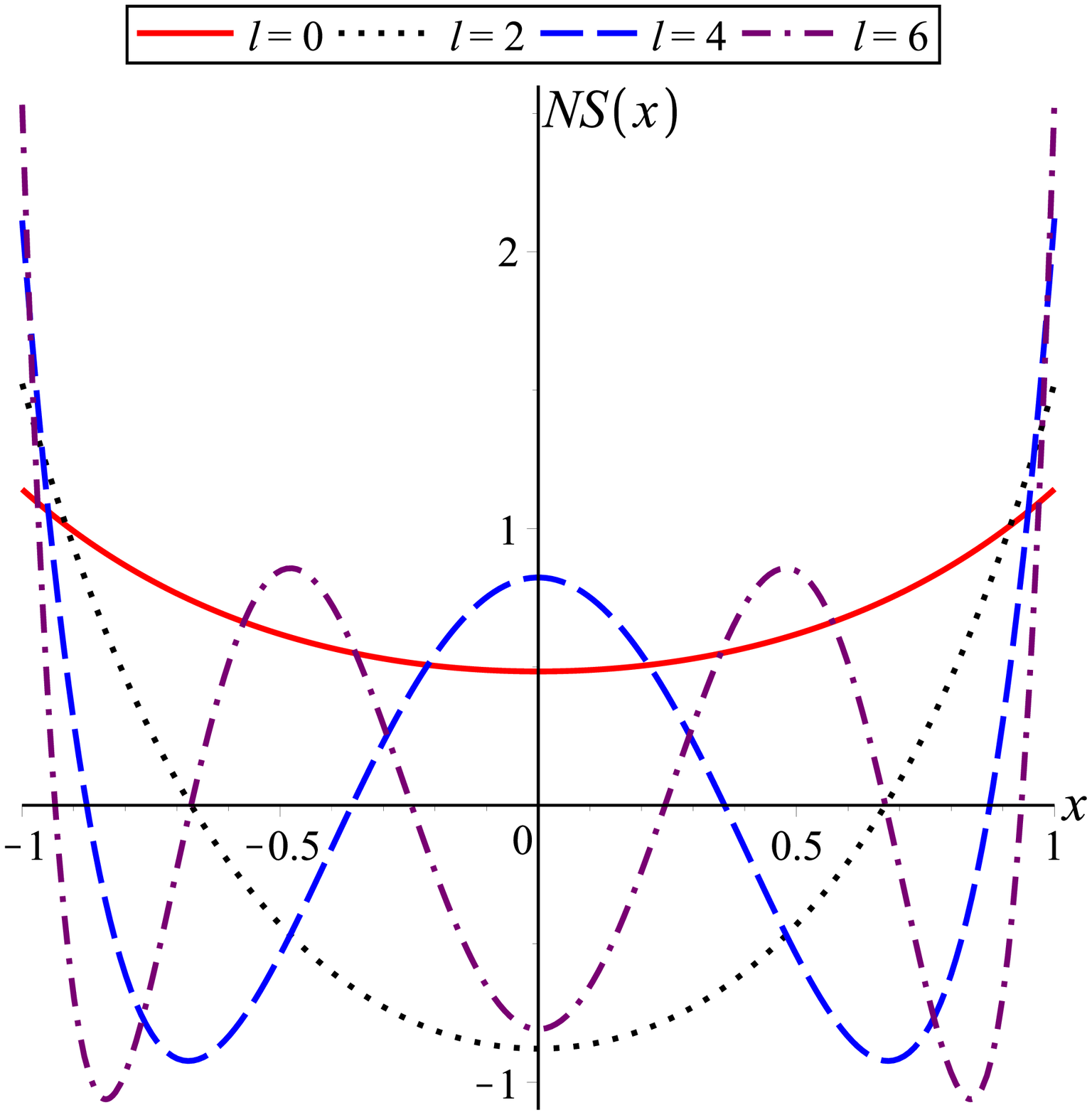}}\label{CA1}\hfil%
\subfigure[]{\includegraphics[width=3.8cm]{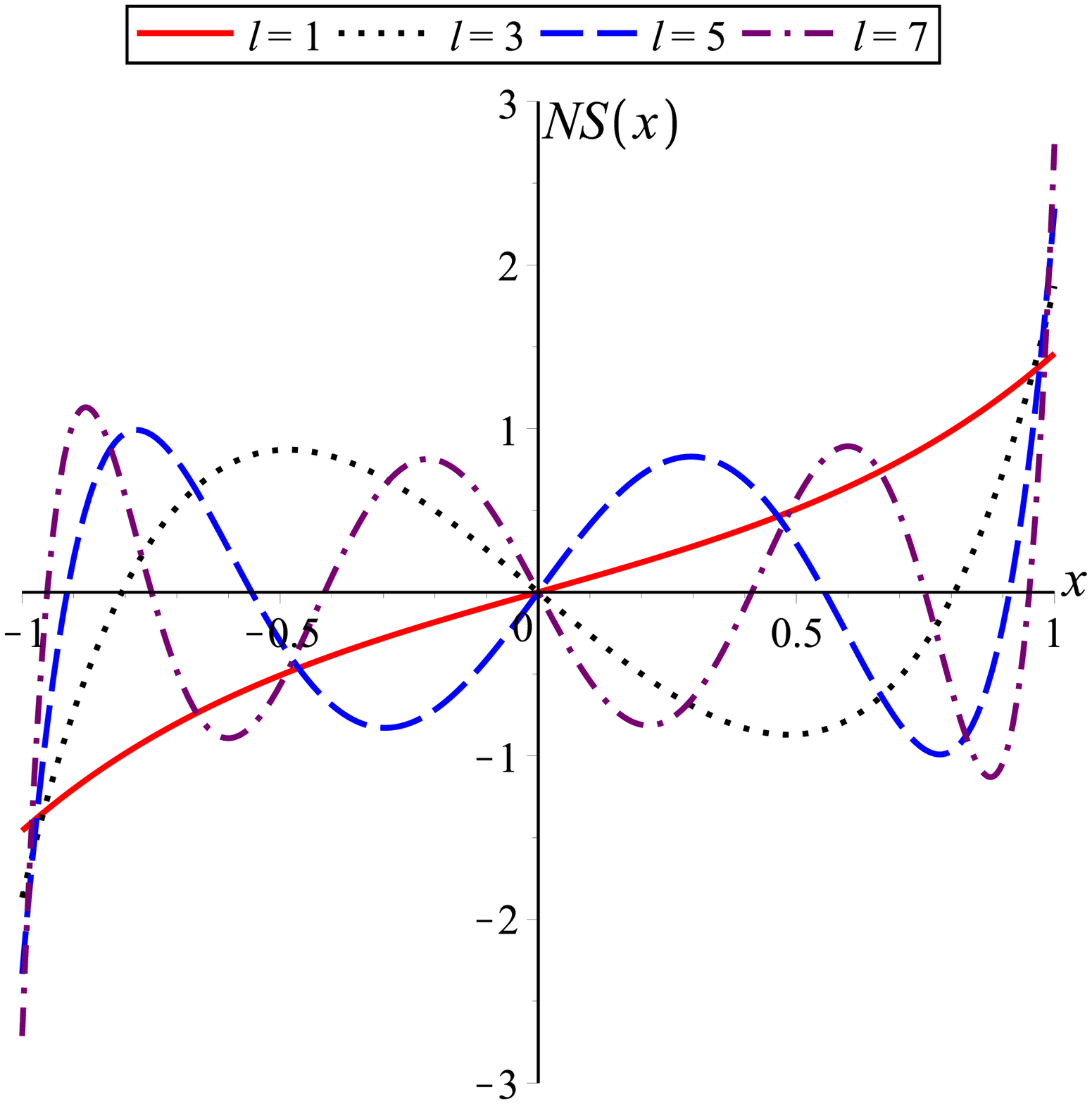}}\label{NA1}\\
\subfigure[]{\includegraphics[width=3.8cm]{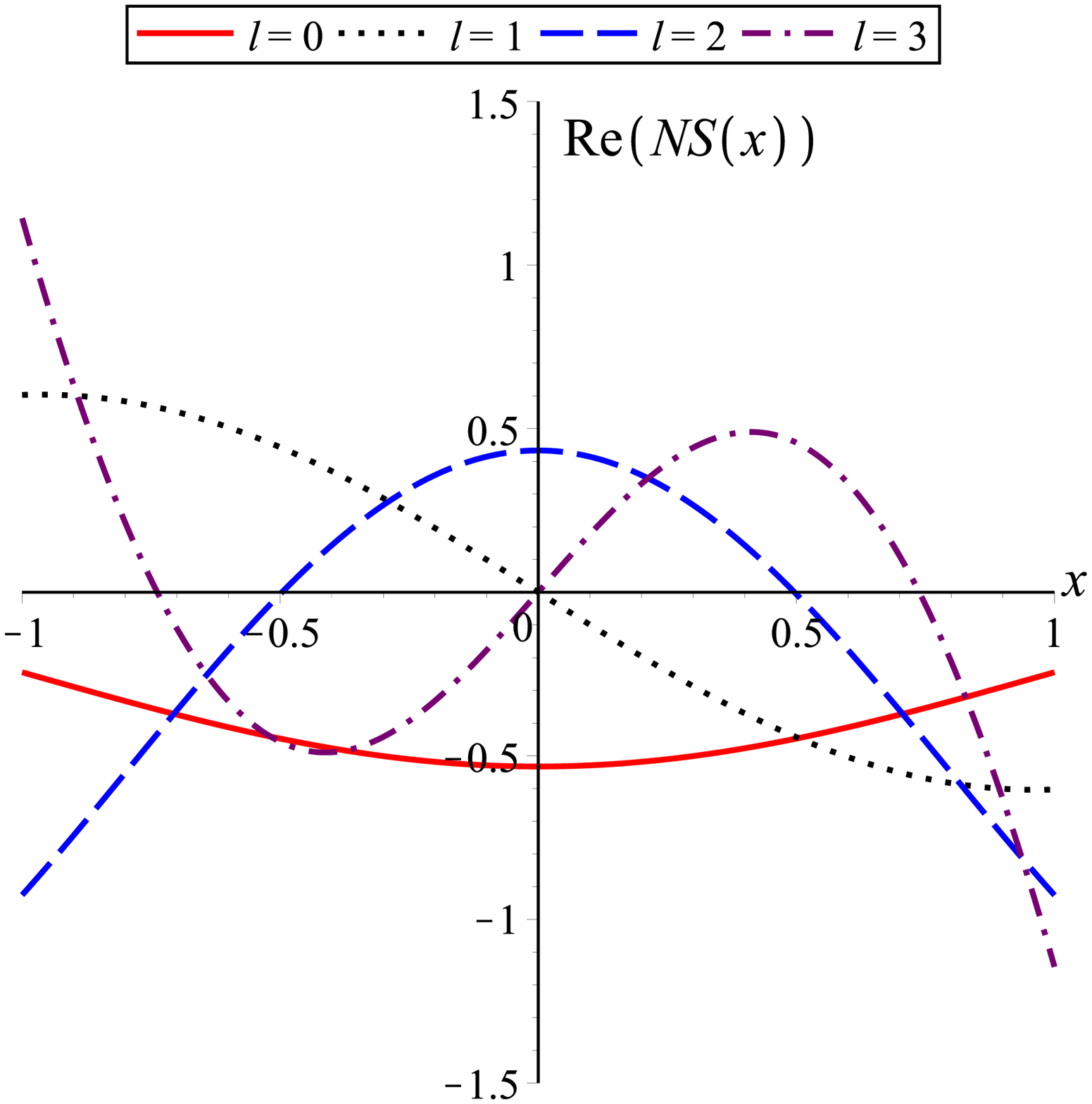}}\label{NA1}
\subfigure[]{\includegraphics[width=3.8cm]{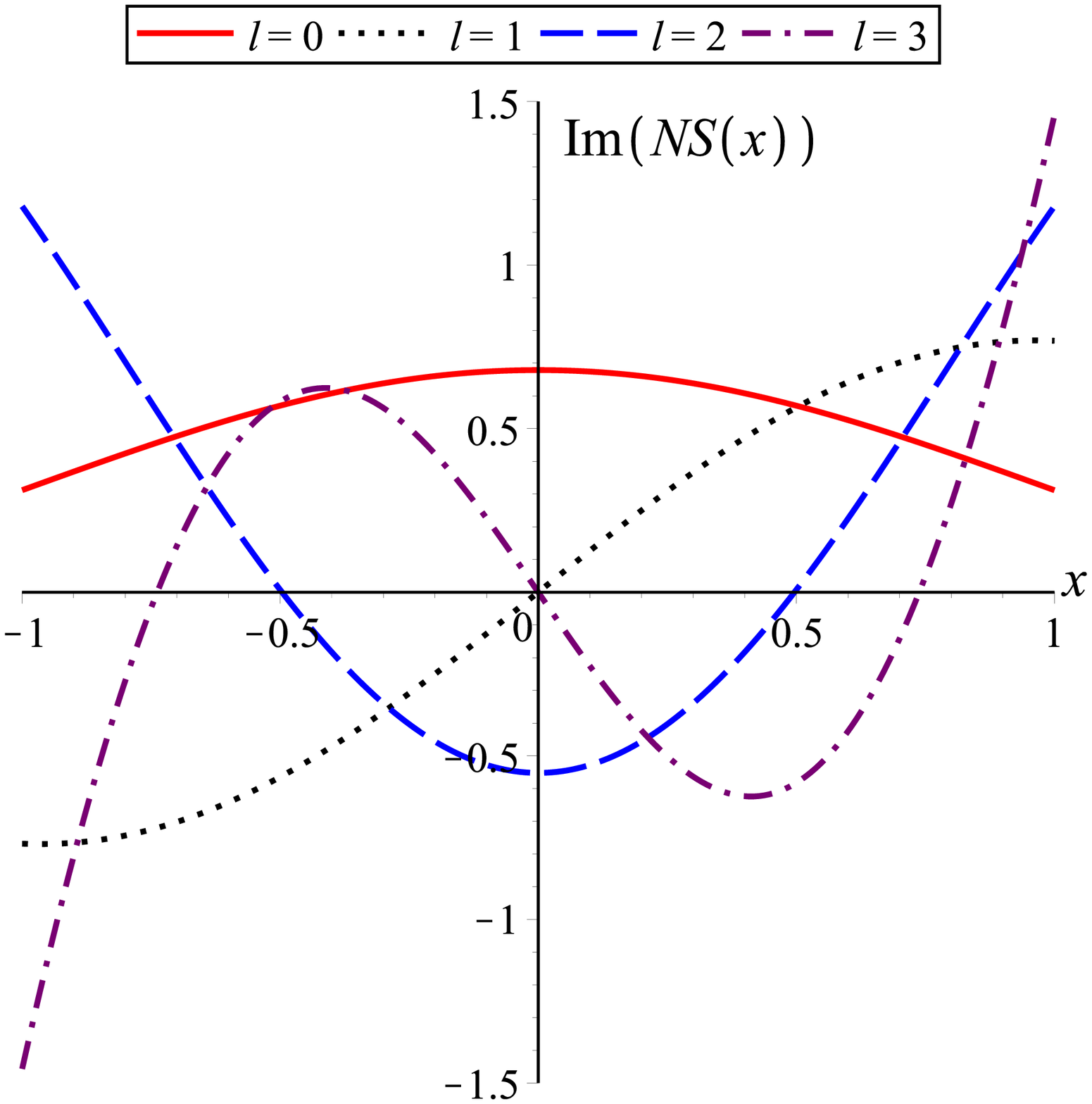}}\label{NA1}
\end{tabular}
\caption{The normalized eigenfunction for real $\tau^2$ (a)	Even parity normalized eigenfunction for $\tau^2 = \;5\;$ (b) Odd parity case, (c) Real part of normalization eigenfunction $\tau^2=-5$
(d) same as (c)  but for imaginary part. }
\label{fig:condiciones}
\end{figure}

\begin{figure}
\begin{tabular}{l}
\subfigure[]{\includegraphics[width=3.8cm]{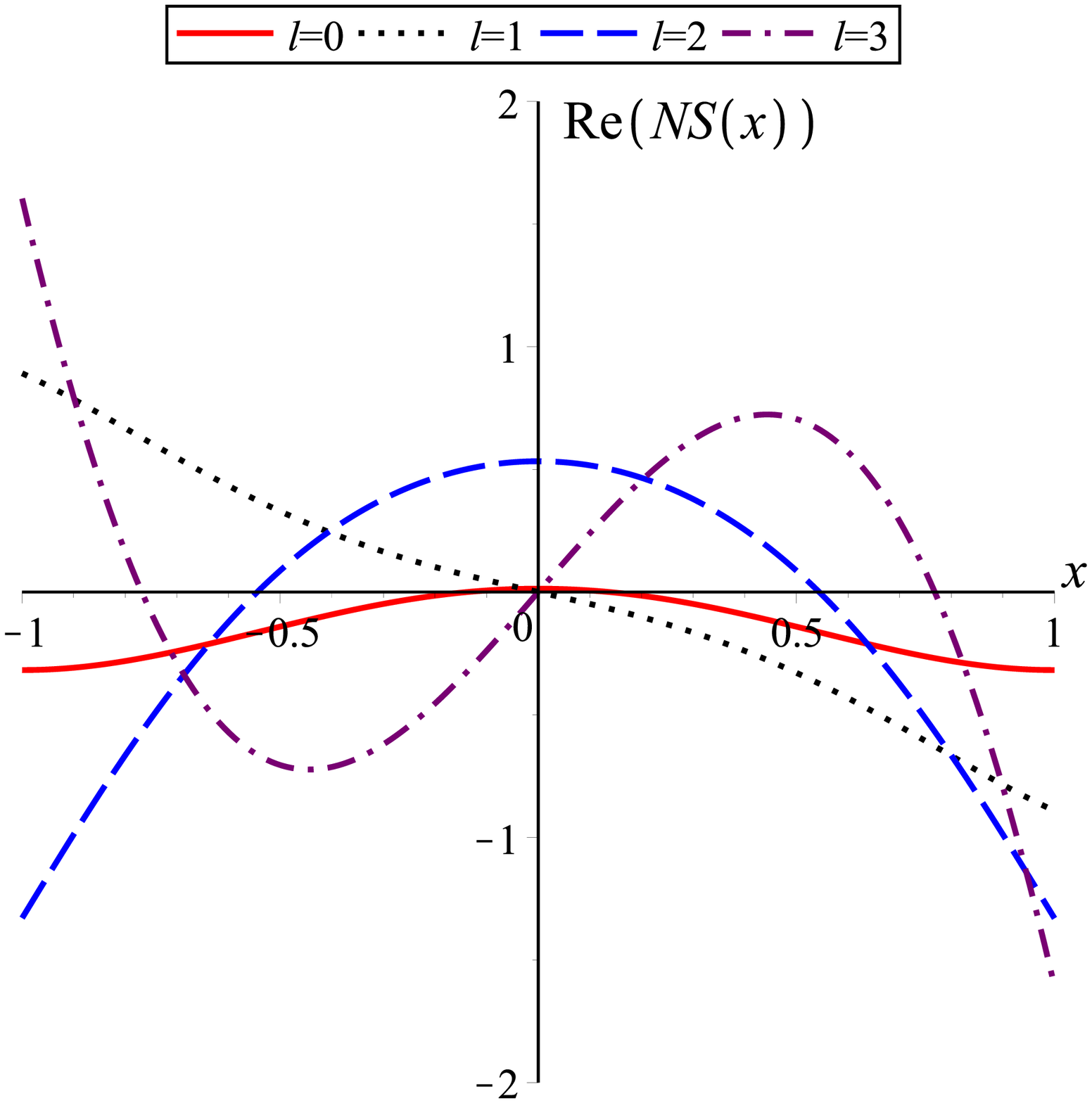}}\label{CA1}\hfil%
\subfigure[]{\includegraphics[width=3.8cm]{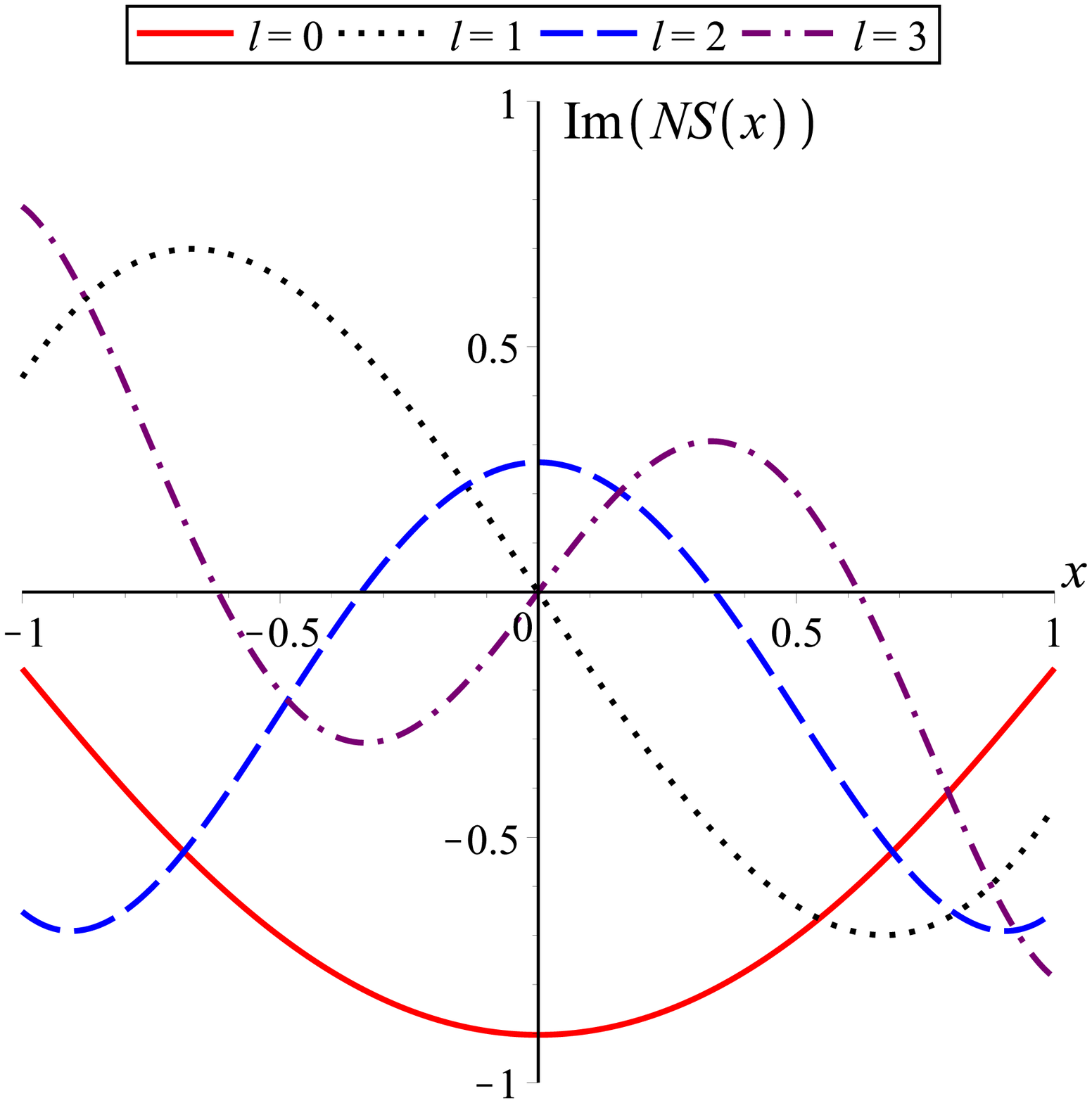}}\label{NA1}
\end{tabular}
\caption{The normalized eigenfucntion for complex $\tau^2=-5-8i$. The real part of normalized eigenfunctions in (a) but imaginary part in (b). }
\label{fig:condiciones}
\end{figure}

Let us illustrate the APD $w(\theta, \phi)=|NY(1)|^2/(2\pi)$ for $x-z$ plane, where $NY(1)$ denotes the normalized eigenfunctions.
For the ground state, the APDs when $\tau^2=-5$ and $\tau^2=\pm 8i$ are displayed in Fig. 4;
the APDs when $\tau^2=-5\pm 8i$ and $\tau^2=5 \pm 8i$ are displayed in Fig. 5.
For different $\tau^2=\tau_R+i\, \tau_I$, we find that the APD of the ground state moves towards the north and south poles for $\tau_R>0$, but gathers to the equator for both $\tau_R\leq 0$.
However, we notice that the APD for large $l$ always moves towards the north and south poles for arbitrary $\tau^2$.

\begin{figure}
\begin{tabular}{l}
\subfigure[]{\includegraphics[width=3.8cm]{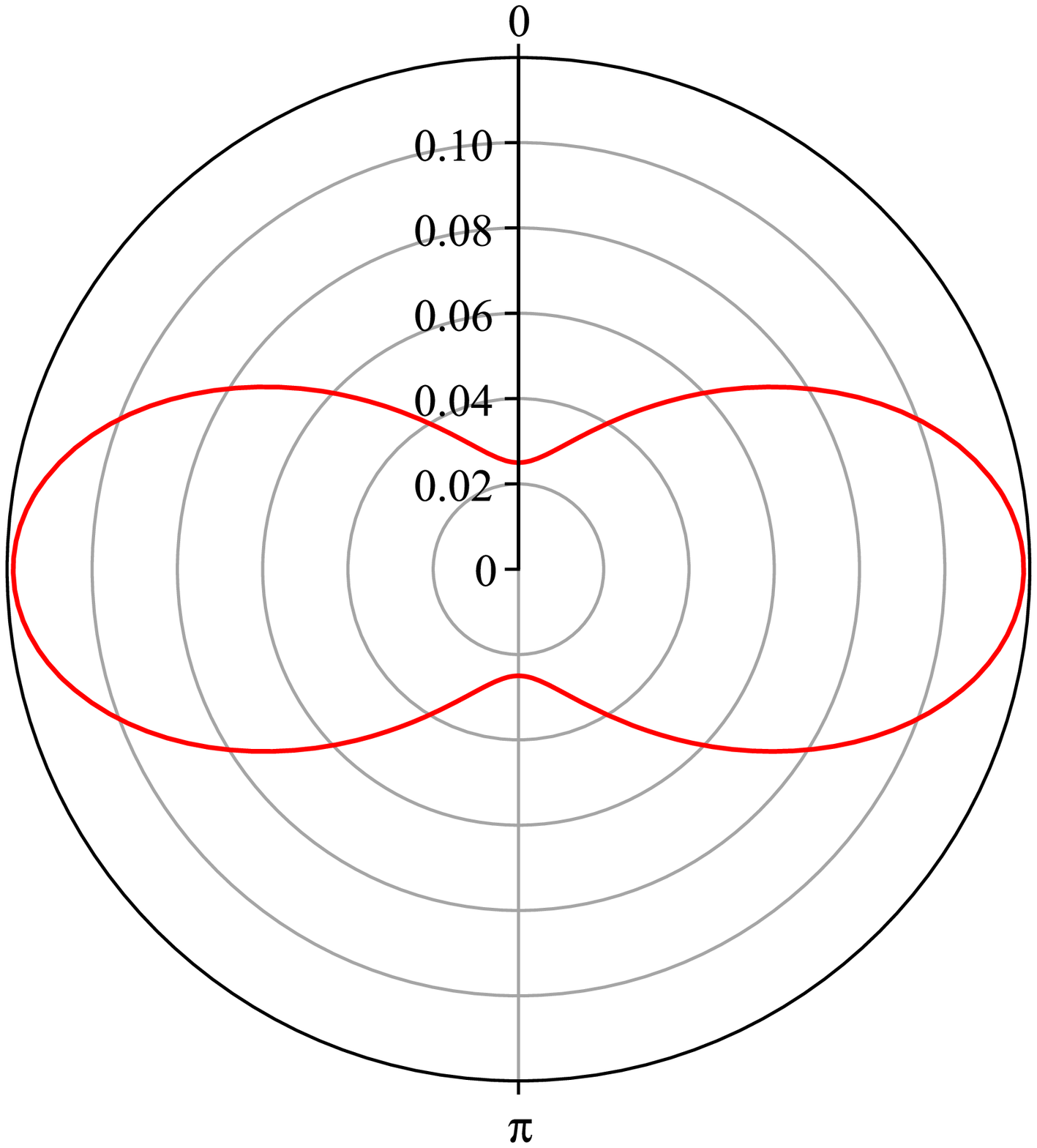}}\label{CA1}\hfil%
\subfigure[]{\includegraphics[width=3.8cm]{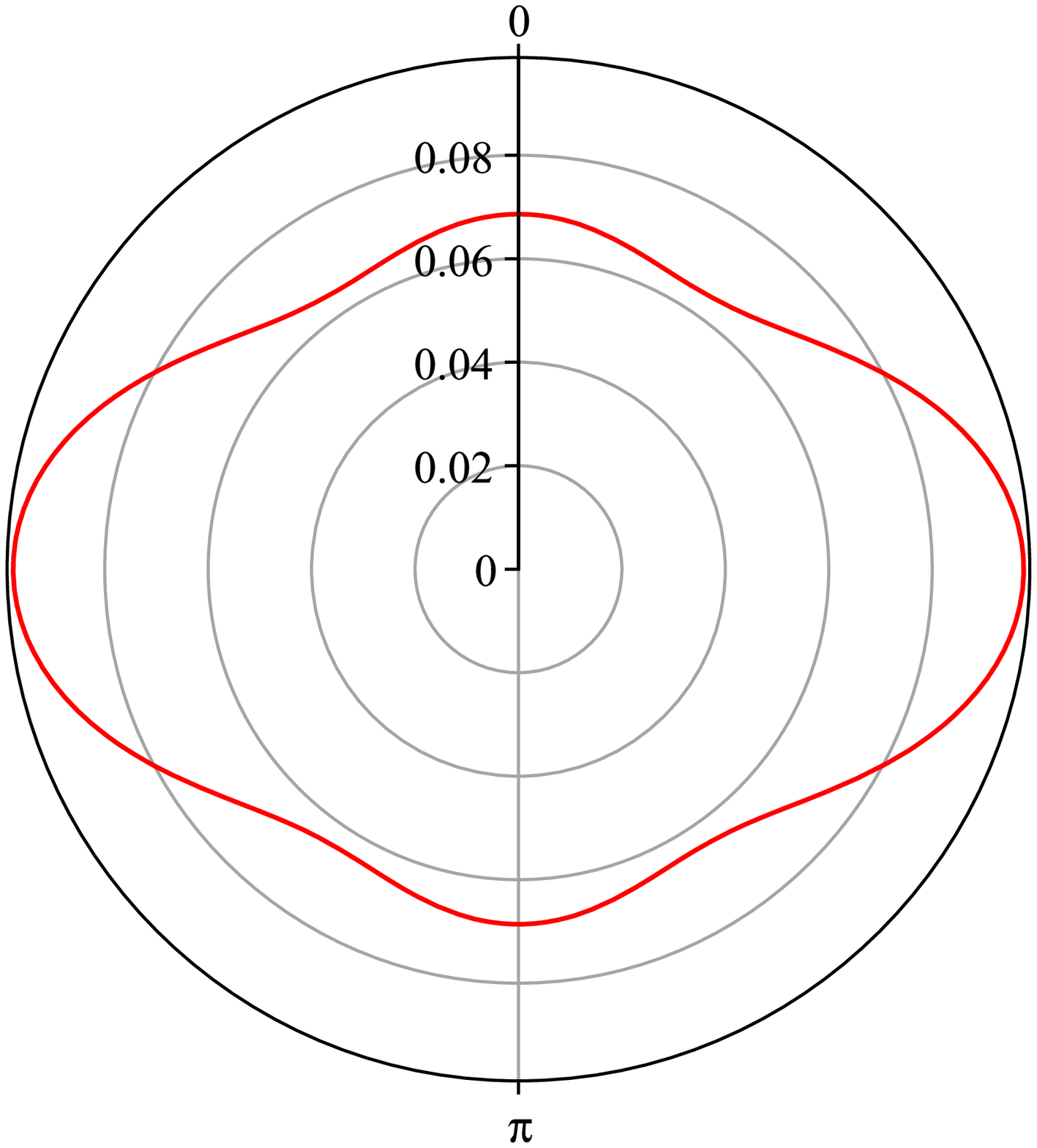}}\label{NA1}\\
\subfigure[]{\includegraphics[width=3.8cm]{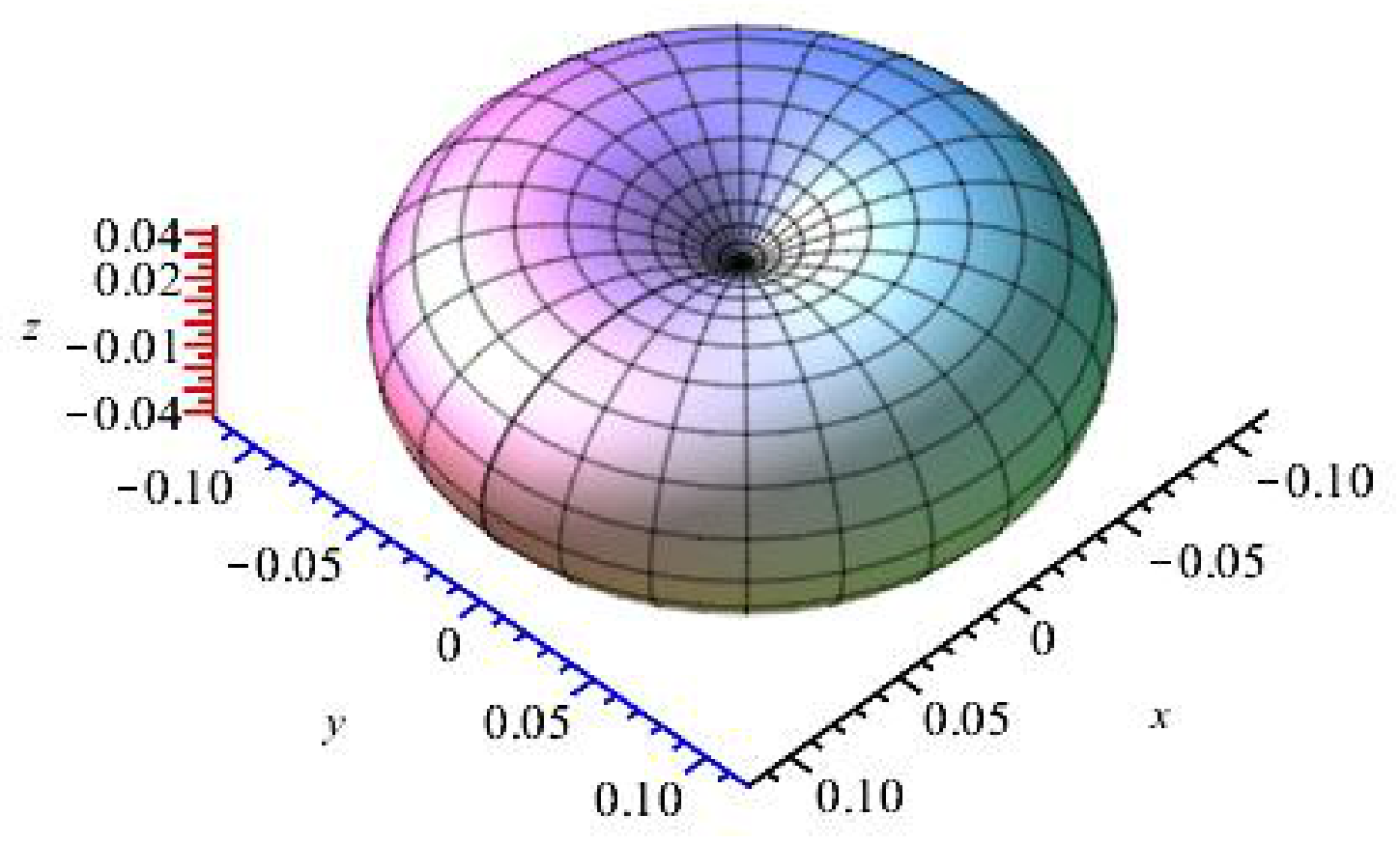}}\label{NA1}
\subfigure[]{\includegraphics[width=3.8cm]{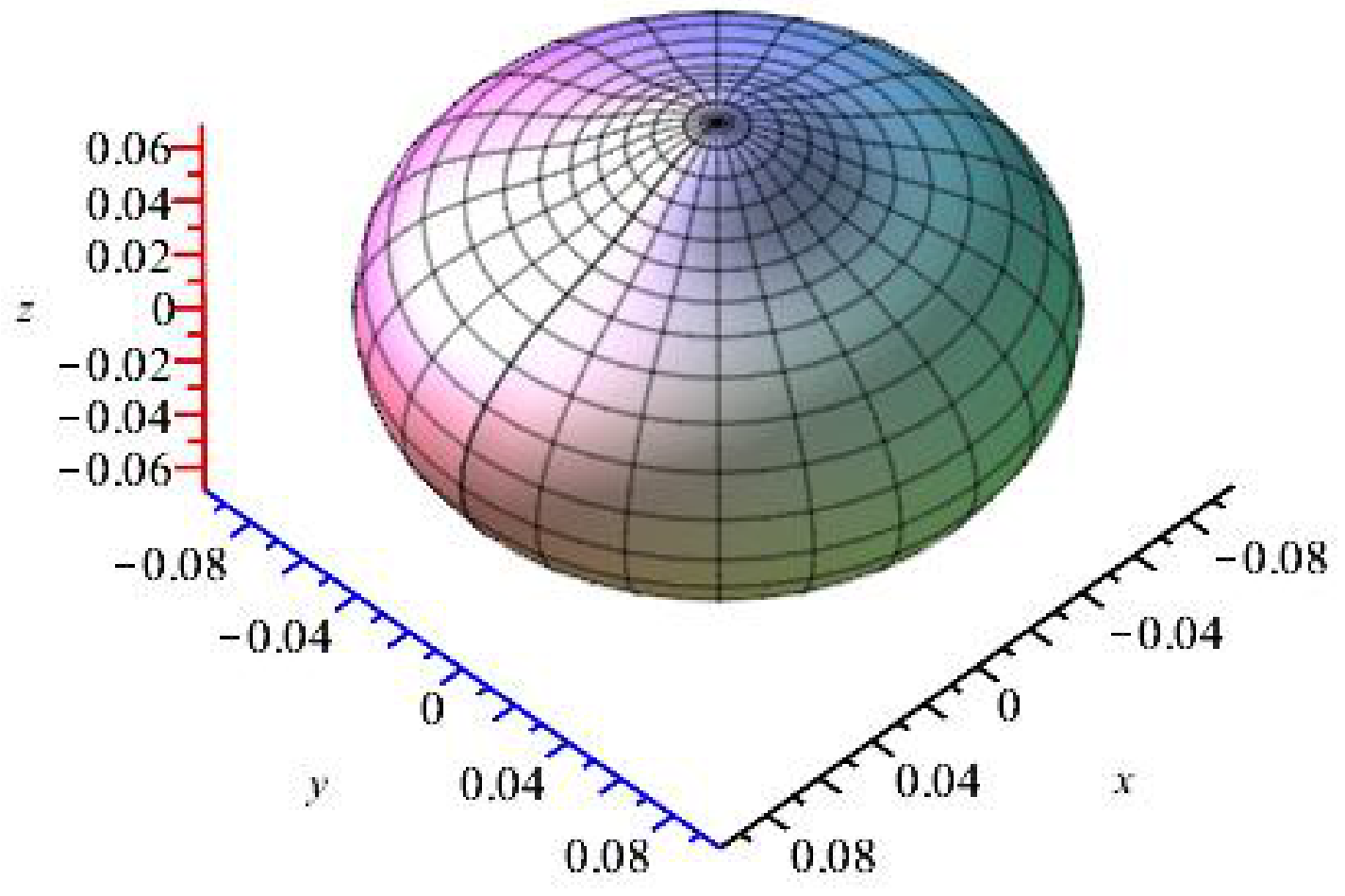}}\label{NA1}
\end{tabular}
\caption{The plots of angular distribution functions in 2D and 3D for the real $\tau^2=-5$ or imaginary number $\tau^2=\pm 8i$. }
\label{fig:condiciones}
\end{figure}

\begin{figure}
\begin{tabular}{l}
\subfigure[]{\includegraphics[width=3.8cm]{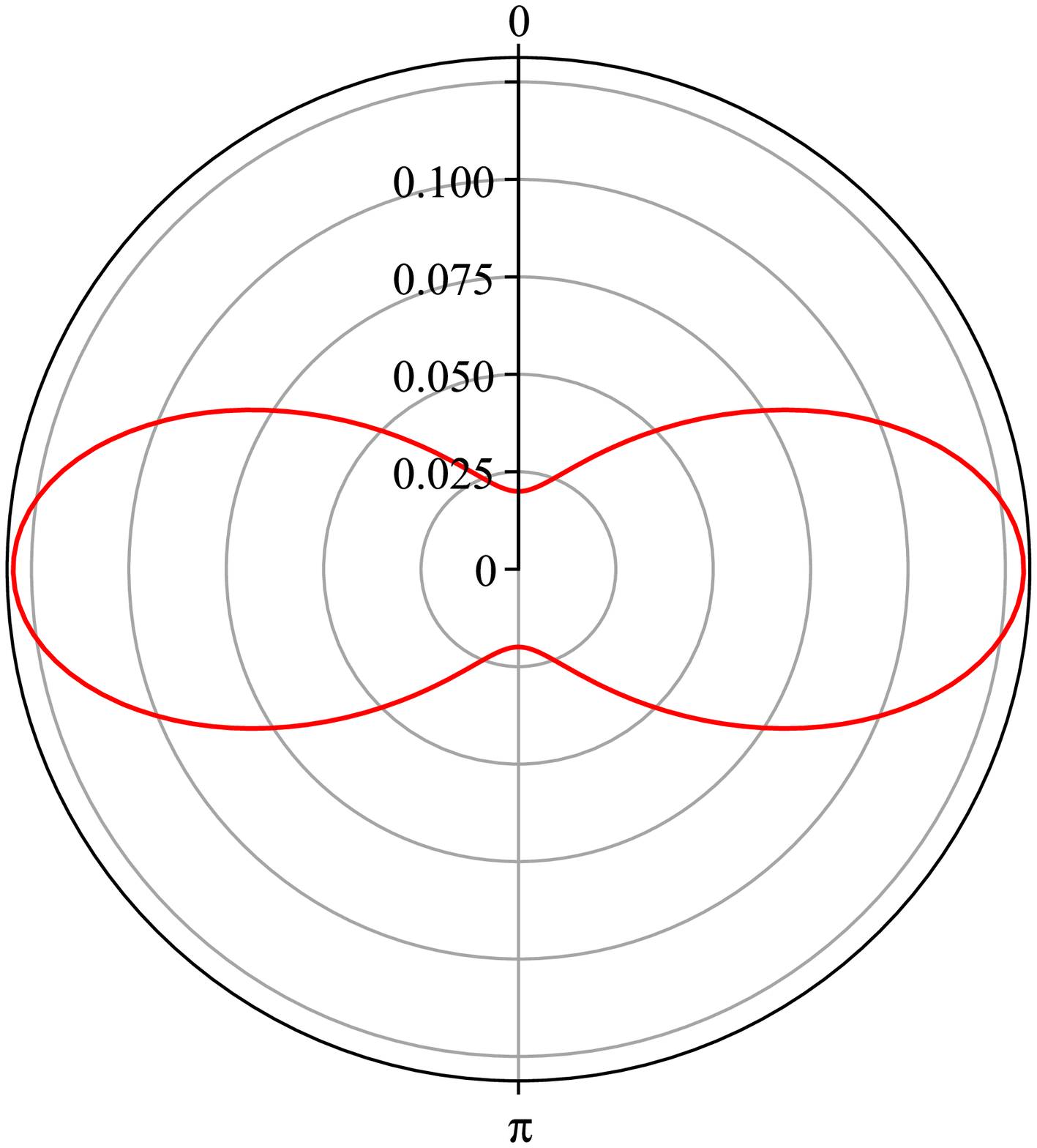}}\label{CA1}\hfil%
\subfigure[]{\includegraphics[width=3.8cm]{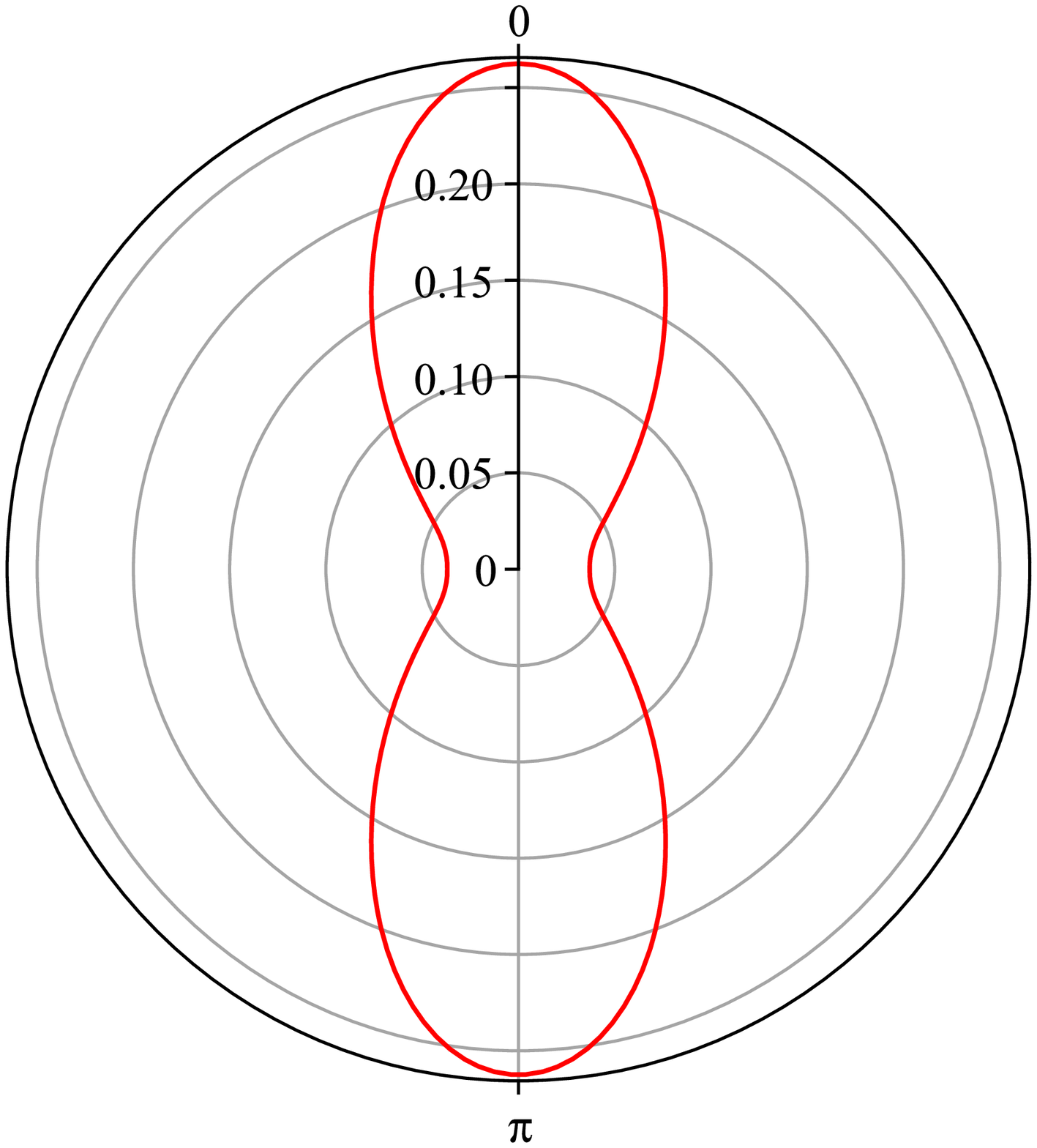}}\label{NA1}\\
\subfigure[]{\includegraphics[width=3.8cm]{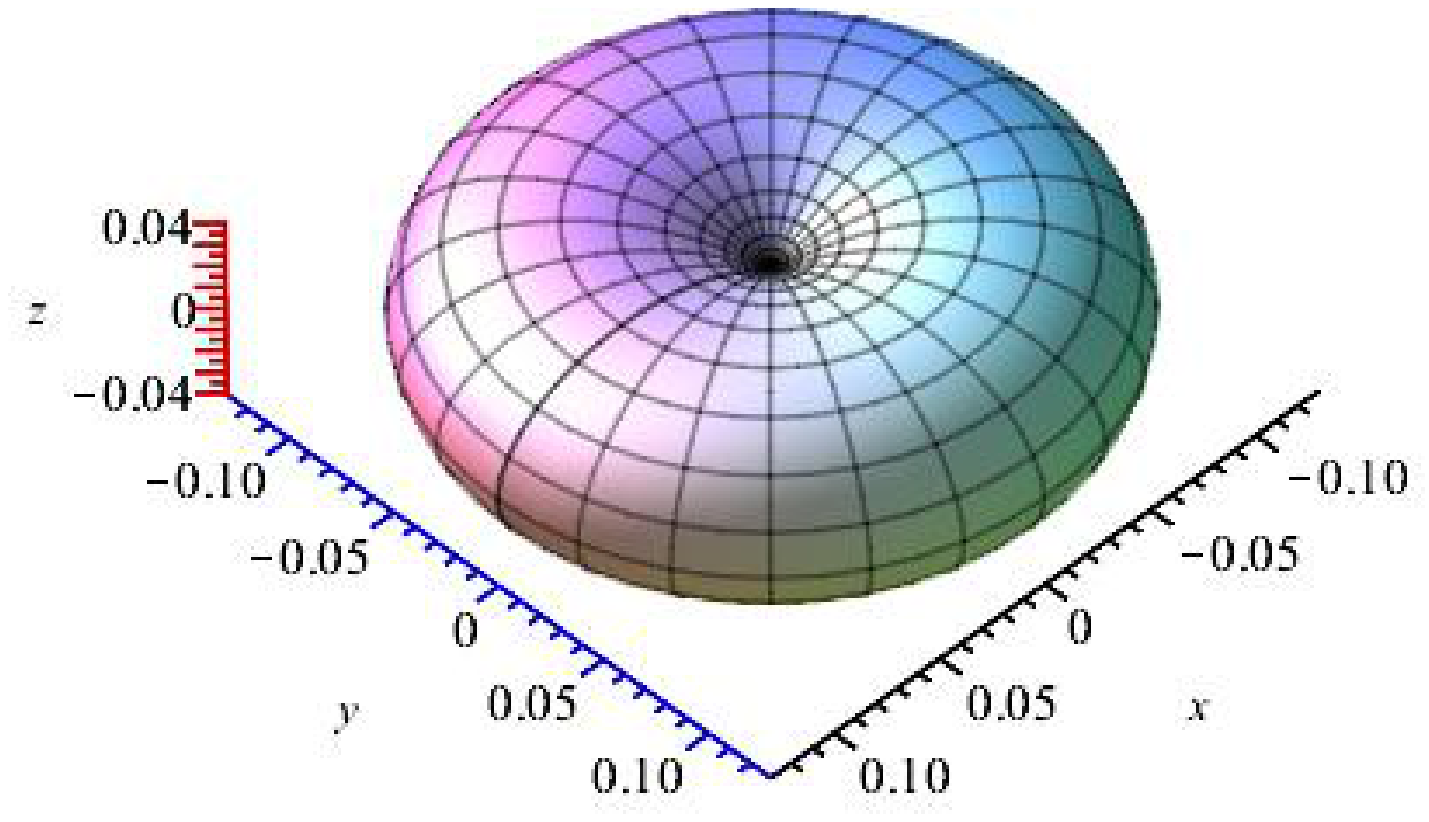}}\label{NA1}
\subfigure[]{\includegraphics[width=3.8cm]{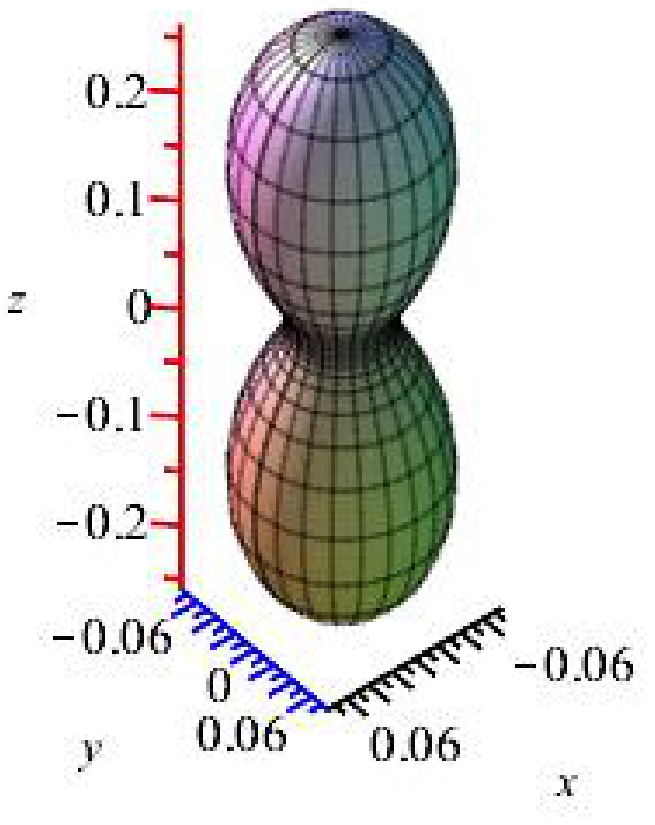}}\label{NA1}
\end{tabular}
\caption{The plots of angular distribution functions in 2D (projection) and 3D for the complex numbers $\tau^2=-5\pm 8i$ and $\tau^2=5\pm 8i$.  }
\label{fig:condiciones}
\end{figure}

\section{Concluding remarks}
In this work, we have proposed a new scheme to solve the angular Teukolsky equation in particular cases. We first transformed this equation to a confluent Heun differential equation via different variable and function transformations and then find two linearly dependent solutions used to constructed the Wronskian determinant (\ref{condition1}). Based on this formula (\ref{condition1}), we are able to calculate the eigenvalues precisely with the aid of Maple. Once eigenvalues were found, we could obtain the normalized eigenfunctions and thus studied the angular probability distribution.

Before ending this work, we are going to give two useful remarks. First, if taking variable transform $z' = 1 - z$ and acting it directly to Eq.(4), equation (4) can also be transformed to Eq.(9). This is a notable feature of the confluent Heun differential equation (5). It tells us that if some physical problem with natural boundary condition can be described by a confluent Heun differential equation, then we are able to obtain its exact solutions following the present scheme. Second, taking ${}_0{S_{l0}}(\tau , x) = {e^{-\tau x}}f(x)$  and then $z = (1+x)/2$ $(x\in[-1, 1], z\in[0, 1])$, Eq. (3) can be transformed to confluent Heun differential equation (4). Similarly, taking ${}_0{S_{l0}}(\tau , x) = {e^{{\rm{ - }}\tau x}}f(x)$  and then choosing $z' = (1-x)/2$ $(x\in[-1, 1], z'\in[0, 1])$, Eq. (3) can be transformed to confluent Heun differential Eq. (9). Following the way discussed above, we obtain the same Wronskian determinant as Eq.(11) and solve for the eigenvalues of Eq. (3). That is to say, the eigenvalues of Eq.(3) are determined totally by the parameter $\tau^2$. The normalized eigenfunctions obtained by Eq.(8) or (10) are linearly dependent within the interval.

\begin{widetext}
\begin{table*}
\caption{Comparison of selected values of eigenvalues computed in Refs. [24, 26] and ours.}
\begin{center}
\begin{tabular}{c | c | c | c | c | c }
\hline
 $l$  & $m$ & $\tau^2(-c^2)$ & ${\rm Yan~} {\it et~ al.}$ [26] & ${\rm Falloon~} {\it et~ al.}$ [24] & ${\rm Present}$\\

\hline
0   &  0 &    -1 &  0.31900 00551 4688 &  ---&   0.31900 00551 46892 73978 39819 9 \\
\hline
0   &  0 &  -100 &  9.22830 42972 498 &  9.22830 42972 49945 15101 22688 &  9.22830 42972 49945 15101 22687 6 \\
1   &  0 &  -100 & --- &   28.13346 37328 26727 81431 89751 & 28.13346 37328 26727 81461 89750 1 \\
\hline
0   &  0 & -2500 &  49.24615 25271 1 & ---&	 49.24615 25271 04644 71397 05257 3 \\
\hline
0   &  0 &  100  &  --- & 	-81.02794 39449 57756 18608 90809	& -81.02794 39449 57756 18608 90808 6  \\
1   &  0 &  100  &  --- &   -81.02793 80237 45584 07315 28426	& -81.02793 80237 45584 07315 28425 3 \\
\hline
\end{tabular}
\end{center}
\end{table*}
\end{widetext}

This work is supported by the National Natural Science Foundation of China under Grant No. 11975196 and partially by project 20190234-SIP-IPN, COFAA-IPN, Mexico and the CONACYT project under grant
No. 288856-CB-2016.

\end{document}